\documentclass[showpacs,showkeys,11pt,
preprint,preprintnumbers,nofootinbib,
groupedaddress,superscriptaddress]{revtex4-1}

\usepackage[dvipdfmx]{graphicx}
\usepackage{graphics}
\usepackage[hypertex]{hyperref}
\usepackage{amsmath,amssymb}	

\begin{document}
\title{Searches for additional Higgs bosons in multi-top-quarks
events at the LHC and the International Linear Collider}
\preprint{UT-HET-100}

\author{Shinya Kanemura}
\email{kanemu@sci.u-toyama.ac.jp}
\affiliation{Department of Physics, University of Toyama, Toyama
930-8555, Japan}
\author{Hiroshi Yokoya}
\email{hyokoya@sci.u-toyama.ac.jp}
\affiliation{Department of Physics, University of Toyama, Toyama
930-8555, Japan}
\affiliation{Theory Center, KEK, Tsukuba 305-0801, Japan}
\author{Ya-Juan Zheng}
\email{yjzheng218@gmail.com}
\affiliation{CTS, CASTS and Department of Physics,
National Taiwan University, Taipei 10617, Taiwan}
\date{\today}

\begin{abstract}
We study direct searches of additional Higgs bosons in multi-top-quarks
 events at the LHC Run-II, its luminosity upgraded version with
 3000~fb$^{-1}$, and the International Linear Collider (ILC) with the
 collision energy of 1~TeV.
Additional Higgs bosons are predicted in all kinds of extended Higgs
 sectors, and their detection at collider experiments is a clear
 signature of the physics beyond the standard model.
We consider two Higgs doublet models with the discrete symmetry as
 benchmark models.
If these additional Higgs bosons are heavy enough, the decay modes
 including top quarks can be dominant, and the searches in
 multi-top-quarks events become an important probe of the Higgs sector.
We evaluate the discovery reach in the parameter space of the model,
and find that there are parameter regions where the searches at the LHC
 with 3000~fb$^{-1}$ cannot survey, but the searches at the ILC 1~TeV
 run can.
The combination of direct searches at the LHC and the ILC is useful to
 explore extended Higgs sectors.
\end{abstract}
\maketitle

\section{Introduction}\label{sec:intro}

A Higgs boson ($h$) was discovered at the LHC Run-I in
2012~\cite{Aad:2012tfa,Chatrchyan:2012ufa}. 
After the discovery, further precision measurements have revealed its
properties, which seem to be quite consistent with those of the Higgs
boson in the standard model (SM) within current experimental
errors~\cite{Aad:2013wqa,Aad:2013xqa,Chatrchyan:2013iaa,%
Chatrchyan:2013mxa}. 
By measuring the mass of the Higgs boson very precisely,
$m_h=125.09\pm0.24$~GeV at the LHC~\cite{Aad:2015zhl}, the parameters in 
the Higgs potential in the SM have been determined experimentally. 
However, the whole structure of the Higgs sector is still unclear, since
one can consider models with extra scalar fields, like the Higgs sector
in the supersymmetric models, satisfying all the available constraints
including the properties of the discovered $h$. 
Moreover, in fact, numbers of models with extra scalar fields have been
proposed to solve the problems which cannot be explained within the SM,
such as neutrino masses, dark matter, baryon asymmetry of the universe,
cosmic inflation, etc.
Determining the structure of the Higgs sector experimentally becomes an
important foothold in constructing the new theory above the electroweak
scale. 

For this purpose, further precision measurements are required to pin down
small deviations from the SM in the properties of
$h$~\cite{Asner:2013psa,Dawson:2013bba}. 
This shall be performed at the LHC Run-II with increased energy and more
accumulated luminosity, and also at the future lepton colliders, such as
the International Linear Collider
(ILC)~\cite{Djouadi:2007ik,Behnke:2013lya}, the Future Circular Collider
(FCC)~\cite{Gomez-Ceballos:2013zzn}, and the Compact LInear Collider 
(CLIC)~\cite{Linssen:2012hp}. 
By measuring deviations from the SM in the properties of $h$, one
can probe an extended Higgs sector with rather large masses of
additional Higgs bosons so that effects by the extended part in the
Higgs sector tend to decouple from the properties of the SM-like
$h$~\cite{Kanemura:2014bqa,Kanemura:2015mxa}.

Another approach to elucidate the whole structure of the Higgs sector is
to directly search for additional Higgs bosons, since their existence
must be a clear evidence of an extended Higgs sector.
Direct searches of additional Higgs bosons have been performed at the
LHC Run-I in various decay modes~\cite{Agashe:2014kda}, and the limits
on the mass of additional Higgs bosons and their coupling strength have
been investigated.
On the prospect of the direct searches at future colliders, the LHC
Run-II has an advantage since it is the energy frontier experiment which
is good to produce heavier particles. 
However, on the other hand, direct searches at the ILC have a different
advantage, that is, a parameter regions with relatively small cross
section can be probed due to the clean environment of lepton collider
experiments, although the mass reach for additional Higgs bosons is
relatively limited. 
Thus, the searches at the LHC Run-II or its luminosity upgraded version
with 3000~fb$^{-1}$ (LHC~3000~fb$^{-1}$) and the ILC can be
complementary to servey the wide parameter regions in extended Higgs
sectors~\cite{Kanemura:2014dea}. 

In this paper, we study multi-top-quarks events as distinct signals
of production of additional Higgs bosons.
As benchmark models of extended Higgs sectors, we consider two
Higgs doublet models (2HDMs).\footnote{%
For the review of the 2HDMs, see, e.g.,
Refs.~\cite{Gunion:1989we,Branco:2011iw}.}
In the 2HDM, there are $CP$-even $H$, $CP$-odd $A$, and a pair of
charged $H^\pm$ in addition to the SM-like $h$.
For neutral $H$ and $A$, the decays into a top-quark pair open if their
masses are larger than about 350~GeV.
Because of the large mass of the top quark, it is quite possible
that this decay mode dominates the branching fraction at least in
certain regions in the parameter space.
Thus, the multi-top-quarks events can be an attractive signal of the
heavy Higgs bosons. 
Such multi-top-quarks events have been studied at the LHC as a signal of
new particle in various models~\cite{Cheung:1995eq,Spira:1997ce,%
Jung:2010ms,Gregoire:2011ka,Cacciapaglia:2011kz,Lillie:2007hd,%
Chen:2008hh,Barger:2010uw,AguilarSaavedra:2011ck,Han:2012qu,Chen:2014ewl,%
Dev:2014yca,Greiner:2014qna,Beck:2015cga,Craig:2015jba}. 
A large production rate of multi-top-quarks events is predicted at the
LHC for the models with colored new particles such as gluinos,
color-octet scalars, etc., but that for non-colored particles such as
heavy additional Higgs bosons is limited.
At the ILC, four top-quarks production can be considered as a signal of
$H$ and $A$, through $e^+e^-\to HA$ and $e^+e^-\to f\bar
fH/A$~\cite{Djouadi:1996ah,Gunion:1988tf,Kanemura:2014dea}. 
To produce a pair of $H$ and $A$ which decay into top-quark pairs, both
the masses of $H$ and $A$ are required to be larger than about 350~GeV,
and the collision energy should be higher than about 700~GeV. 
Such an experiment can be realized at the ILC with $\sqrt{s}=1$~TeV. 
Up to our knowledge, there has been no dedicated study on this process
at lepton colliders. 
Therefore, in this paper, we aim to present a detailed analysis on this
process including the hadron-level simulation with jet clustering,
flavor tagging, detector acceptance and momentum resolution effects.
We find that the four top-quarks events can be detected by simple
kinematical cuts, and thus be useful to survey the parameter regions in
the 2HDM. 
We note that the signal of heavy charged Higgs boson can be $H^\pm\to
t\bar b(\bar t b)$, and its observability has been studied in $gb\to
tH^\pm$ process at hadron
colliders~\cite{Borzumati:1999th,Plehn:2002vy}, and also at lepton
colliders in $e^+e^-\to H^+H^-$ and $e^+e^-\to H^\pm
f\bar{f}'$~\cite{Kanemura:2000cw,Moretti:2002pa,Kiyoura:2003tg}. 

The paper is organized as follows.
In Sec.~\ref{sec:2hdm}, we briefly introduce the 2HDM with $Z_2$
symmetry considering the four types of Yukawa interactions.
In Sec.~\ref{sec:lhc}, we present an analysis for the search prospect of
additional Higgs bosons in multi-top-quarks events at the LHC.
In Sec.~\ref{sec:ilc}, we study the four top-quarks events at the
ILC by performing the Monte-Carlo simulation for the signal and
background processes at the hadron level with detector effects.
By using the simulation analysis, we evaluate the discovery potential of
the neutral Higgs bosons at the ILC in the four top-quarks events in the
parameter space in the 2HDM with four types of Yukawa interactions.
The obtained discovery reaches at the LHC and at the ILC are compared.
Sec.~\ref{sec:dis} is devoted to discussions for further investigation
and future prospects.
Finally, we draw a conclusion in Sec.~\ref{sec:conc}.

\section{Two Higgs Doublet Model}\label{sec:2hdm}

In this section, we briefly introduce the model we consider, namely the
2HDM with the softly-broken $Z_2$ symmetry.
We introduce two isospin doublet scalar fields, $\Phi_1$ and $\Phi_2$,
which transform as $\Phi_1\to+\Phi_1$ and $\Phi_2\to-\Phi_2$ under the
$Z_2$ transformation.
For the SM fermions, there are four kinds of $Z_2$ parity
assignment~\cite{Barger:1989fj,Grossman:1994jb,Aoki:2009ha}, as listed
in Table~\ref{tab:Z2}. 
We denote the four types of Yukawa interactions as Type-I, Type-II, Type-X
and Type-Y~\cite{Aoki:2009ha}.
The Yukawa interaction to the SM fermions in each flavor is allowed
for only one Higgs field, $\Phi_1$ or $\Phi_2$, to make each interaction
term $Z_2$ invariant.
It forbids the flavor changing neutral currents at the
tree-level~\cite{Glashow:1976nt}, which are severely constrained by
experimental observations. 

\begin{table}[t]
 \begin{tabular}{c|ccccccc}
  \hline
  & $\Phi_1$ & $\Phi_2$ & $u_R$ & $d_R$ & $\ell_R$ &
  $Q_L$ & $L_L$ \\
  \hline
  \hline
  Type-I & $+$ & $-$ & $-$
              & $-$ & $-$ & $+$
                          & $+$  \\
  Type-II & $+ $& $-$ & $-$
  & $+$ & $+$ & $+$
  & + \\
  Type-X & $+$ & $-$ & $-$
  & $-$ & $+$ & $+$
  & $+$  \\
  Type-Y & $+$ & $-$ & $-$
  & $+$ & $-$ & $+$
  & $+$  \\
  \hline
 \end{tabular}
\caption{Four possible $Z_2$ charge assignments to the scalar and fermion
 fields.}
\label{tab:Z2}
\end{table}

After solving the condition for the electroweak symmetry breaking and
diagonalizing the mass matrices in the Higgs sector assuming
$CP$-invariance, there are five physics scalar states; namely, two
$CP$-even neutral Higgs boson $h$ and $H$, one $CP$-odd neutral Higgs
boson $A$, and a pair of charged Higgs bosons $H^\pm$.
The lighter $CP$-even $h$ can be identified as the SM-like Higgs boson
which was observed at the LHC.
Yukawa interactions are given in terms of these physical scalars as 
\begin{align}
 -{\cal L}_{\rm Yukawa} &= \sum_{f=u,d,\ell}\left[
\frac{m_f}{v}\xi_h^f\bar{f}fh + \frac{m_f}{v}\xi_H^f\bar{f}fH
-i\frac{m_f}{v}\xi_A^f\bar{f}\gamma_5fA \right]\nonumber\\
& + \left\{
\frac{\sqrt{2}V_{ud}}{v}\bar{u}\left[m_u\xi_A^u{\rm P}_L
+m_d\xi_A^d{\rm P}_R\right]dH^+ +
 \frac{\sqrt{2}m_\ell}{v}\xi_A^\ell\overline{\nu_L}\ell_RH^+ + {\rm H.c.}
\right\},
\end{align}
where the scaling factor $\xi_\phi^f$ with $\phi=h,H,A$ and $f=u,d,\ell$
can be found in Table~\ref{tab:yukawa}.
The scaling factor is a function of $\alpha$ and $\beta$, the mixing
angles in the neutral $CP$-even component and $CP$-odd component, 
respectively. 
The mixing angle $\beta$ also satisfies $\tan\beta=v_2/v_1$, where $v_1$
and $v_2$ are the vacuum expectation values of the two doublet fields. 

\begin{table}[t]
 \begin{tabular}{c|ccccccccc}
  \hline
  & $\xi_h^u$ & $\xi_h^d$ & $\xi_h^\ell$ & $\xi_H^u$ & $\xi_H^d$ &
  $\xi_H^\ell$ & $\xi_A^u$ & $\xi_A^d$ & $\xi_A^\ell$ \\
  \hline
  \hline
  Type-I & $c_\alpha/s_\beta$ & $c_\alpha/s_\beta$ & $c_\alpha/s_\beta$
              & $s_\alpha/s_\beta$ & $s_\alpha/s_\beta$ & $s_\alpha/s_\beta$
                          & $\cot\beta$ & $-\cot\beta$& $-\cot\beta$ \\
  Type-II & $c_\alpha/s_\beta$ & $-s_\alpha/c_\beta$ & $-s_\alpha/c_\beta$
  & $s_\alpha/s_\beta$ & $c_\alpha/c_\beta$ & $c_\alpha/c_\beta$
  & $\cot\beta$ & $\tan\beta$& $\tan\beta$ \\
  Type-X & $c_\alpha/s_\beta$ & $c_\alpha/s_\beta$ & $-s_\alpha/c_\beta$
  & $s_\alpha/s_\beta$ & $s_\alpha/s_\beta$ & $c_\alpha/c_\beta$
  & $\cot\beta$ & $-\cot\beta$& $\tan\beta$ \\
  Type-Y & $c_\alpha/s_\beta$ & $-s_\alpha/c_\beta$ & $c_\alpha/s_\beta$
  & $s_\alpha/s_\beta$ & $c_\alpha/c_\beta$ & $s_\alpha/s_\beta$
  & $\cot\beta$ & $\tan\beta$& $-\cot\beta$ \\
  \hline
\end{tabular}
\caption{The scaling factors $\xi_\phi^f$ for the four types of Yukawa
 interactions~\cite{Aoki:2009ha}.
$c_\theta=\cos\theta,~{\rm and }~s_\theta=\sin\theta$ for $\theta = 
\alpha,~\beta$.}
\label{tab:yukawa}
\end{table}

The gauge coupling of $h$ is given by $g_{hVV}^{2HDM}=g_{hVV}^{\rm
SM}\sin(\beta-\alpha)$ and that of $H$ is given by
$g_{HVV}^{2HDM}=g_{hVV}^{\rm SM}\cos(\beta-\alpha)$. 
Theoretically, a deviation of $\sin(\beta-\alpha)$ from unity is
constrained by the arguments of perturbative
unitarity~\cite{Kanemura:1993hm,Akeroyd:2000wc,Ginzburg:2005dt} and
vacuum stability~\cite{Deshpande:1977rw,Kanemura:1999xf,Nie:1998yn}. 
If a soft-breaking scale of the $Z_2$ symmetry $M$ is larger than the
electroweak scale, $M\gg v$, only small value of $1-\sin(\beta-\alpha)$
is allowed by these constraints~\cite{Gunion:2002zf}.
The limit of $\sin(\beta-\alpha)\to1$ is called the SM-like limit, where
$h$ has the same coupling constants to the gauge bosons and also to the
SM fermions as the SM Higgs boson.
On the other hand, Yukawa interactions of $H$, $A$ and $H^\pm$ to the SM
fermions do not vanish in this limit, and the coupling strength for each
vertex depends on the type of Yukawa interactions and $\tan\beta$. 
Thus, the variety of the type of Yukawa interactions with different
$\tan\beta$ dependences leads to rich phenomenology for the additional
Higgs bosons. 

We focus on the interactions of additional Higgs bosons to top
quarks.
For any type of Yukawa interactions, the Yukawa coupling constants to
top quarks are enhanced by $\cot\beta$ for small $\tan\beta$ regions.
Therefore, for larger masses of additional Higgs bosons where the
decay modes into top quarks are open, the branching ratio to top quarks
becomes dominant for small $\tan\beta$ regions.
The figures for the branching ratio of additional Higgs bosons can be
found, e.g., in Refs.~\cite{Kanemura:2014bqa,Kanemura:2014dea}.
For larger $\tan\beta$, the dominant branching ratio is replaced by the
other fermionic mode, $b\bar b$ for Type-II and Type-Y, $\tau^+\tau^-$
for Type-X Yukawa interactions.
For Type-I, since the $\tan\beta$ dependence is common for all fermions,
the dominance of $t\bar t$ decay mode is true for any value of
$\tan\beta$.

\section{multi-top-quarks production at the LHC}\label{sec:lhc}

In this section, we study the four top-quarks production through the
production of additional Higgs boson(s) at the LHC.
The largest contribution comes from the top-quark pair associated
production process, 
\begin{align}
 pp\to t\bar t H(t\bar t A)\to t\bar t t\bar t, 
\end{align}
since it emerges via the strong interaction.
At the tree level, there are two subprocesses in this process.
One is $gg\to t\bar t H(t\bar t A)$ and the other is $q\bar q \to t\bar
t H(t\bar t A)$.
Because $H(A)$ is radiated off from the top quarks, the cross section is
proportional to the square of $y^{H}_t$ ($y_{t}^A$) which is
proportional to $\cot\beta$ in the SM-like limit.
Therefore, the cross section is large for smaller $\tan\beta$.
 
The four top-quarks production through the pair production of $H$ and $A$, 
\begin{align}
 pp\to HA\to t\bar t t\bar t,
\end{align}
 is described by the quark anti-quark annihilation process, $q\bar q \to
 Z^* \to HA$ at the tree level.
Since the $HA$ production cross section does not depend on $\tan\beta$,
 the cross section of the final four top-quarks production depends on
 $\tan\beta$ through the branching ratios of $H$ and $A$ into the
 top-quark pair, ${\mathcal B}^{H/A}(t\bar t)$.

There are also three top-quarks production processes via the associated
production of $H^\pm$ and $H$(A), which subsequently decay into $t\bar
b$($\bar t b$) and $t\bar t$, respectively, 
\begin{align}
 pp\to H^\pm H (H^\pm A) \to t\bar b t\bar t /\bar t b t\bar t.
\end{align}
At the tree level, this production process is described by the
$W$ boson mediated diagram~\cite{Kanemura:2001hz,Cao:2003tr}, 
\begin{align}
 q\bar q' \to W^{*}\to H^\pm H (H^\pm A).
\end{align}

\begin{figure}[t]
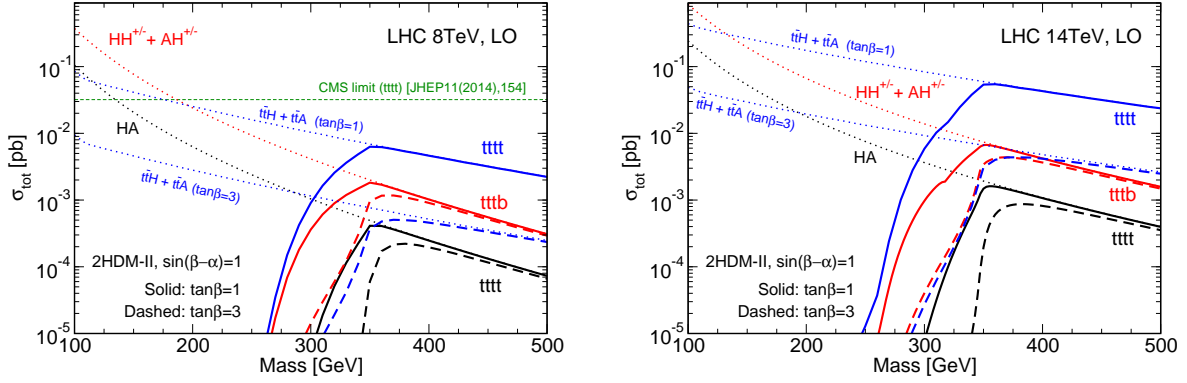

 \begin{center}
  \includegraphics[width=.45\textwidth,clip]{LHC8_HA_multitop.eps}
  \hspace{.5cm}
  \includegraphics[width=.45\textwidth,clip]{LHC14_HA_multitop.eps}
 \end{center}
 \caption{Cross sections for multi-top-quarks production at the LHC
 8~TeV (left) and 14~TeV (right) for the Type-II 2HDM.
four top-quarks production from $t\bar tH+ t\bar tA$ production, 
$HA$ production and three top-quarks production from $HH^\pm+AH^\pm$
 production are shown with $\tan\beta=1$ (solid lines) and 3
 (dashed lines). 
}\label{fig:LHC_HA}
\end{figure}

We estimate the cross sections for these processes at the LHC at
the tree level.
As an example, we take the 2HDM with Type-II Yukawa interactions.
For simplicity, we take a common mass for all the additional Higgs
bosons.
Our calculation is performed with the use of analytic equations in
Ref.~\cite{Djouadi} and the numerical codes generated by {\tt
MadGraph5}~\cite{Alwall:2011uj}.
To calculate the branching fractions of $H\to t\bar t$ and $A\to t\bar
t$, the off-shell effect of top quarks is included.
To estimate the hadronic cross section, the {\tt CTEQ6L} parton
distribution functions (PDFs)~\cite{CTEQ6L} are used with setting the
scale of PDFs to $\mu_F=m_{\Phi_1}+m_{\Phi_2}$ for $HA$ and $H^\pm
H+H^\pm A$ production processes, and to $\mu_F=m_{t}+m_{\Phi}/2$ for
the $t\bar tH(A)$ production process~\cite{Djouadi}.
In Fig.~\ref{fig:LHC_HA}, the results are plotted as a function of the
mass for the LHC 8~TeV (left) and the LHC 14~TeV (right). 
The cross sections for $t\bar t H+t\bar t A$, $HA$ and $H^\pm H+H^\pm A$
productions are plotted in blue, black and red dotted lines,
respectively.
For the $t\bar t H+t\bar t A$ production process, the results for
$\tan\beta=1$ and 3 are plotted. 
In addition, the resulting cross sections for the multi-top-quark
production processes are plotted in solid (dashed) lines with the same
color, assuming that $\tan\beta=1$ and 3.
The cross sections at 13~TeV collision energy are about 70-80\%
of those at 14~TeV. 

For each process, the largest cross section is realized for the mass
of additional Higgs bosons at around 350~GeV.
Below this value, the branching ratio into $t\bar t$ is suppressed
because one of the top quarks is forced to be off-shell, while above
that value the production cross sections of additional Higgs bosons get
decreased. 
For $\tan\beta=1$, the cross section of the four top-quarks production
can be at most 6~fb (50~fb) for the LHC 8~TeV (14~TeV).
We note that there has been already an experimental upper limit for the
cross section of four top-quarks production at the LHC
8~TeV~\cite{TheATLAScollaboration:2013jha,Khachatryan:2014sca}. 
The CMS Collaboration has set the limit to $\sigma_{4t}\le32$~fb at the
95\% CL (confidence level)~\cite{Khachatryan:2014sca}, by observing the
lepton plus multi-jets events.
However, the limit is not tight enough to constrain the parameter
regions in the 2HDM. 
We also note that the SM prediction to the four top-quarks production at
the LHC 8~TeV is about 1~fb~\cite{Barger:2010uw}.
For the three top-quarks production, since the expected cross section is
smaller than that of four top-quarks production while the signatures mix
up with that of four top-quarks production, the detection may be more
challenging. 

We now study the prospect of measuring four top-quarks production as a
signal of the production of additional Higgs bosons at the future LHC
run. 
For the four top-quarks production within the SM contribution, the
total cross section and its uncertainty are estimated to be $\sigma_{\rm
SM}=15$~fb and $\delta\sigma_{\rm SM}=4$~fb,
respectively~\cite{Bevilacqua:2012em}. 
To deal with the background processes, we follow the analysis in
Ref.~\cite{Lillie:2007hd} where the selection cuts to extract the four
top-quarks events out of the background events are demonstrated by
simulation analysis.
In their analysis, the background rate of $B=7.2$~fb after selection
cuts is obtained with the signal efficiency of $\epsilon=0.03$.
By taking into account the statistical and systematical
uncertainties for the signal, SM and background processes, the
accuracy of measuring the signal cross section $\sigma_S$ can be
estimated as 
\begin{align}
 \frac{\delta \sigma_S}{\sigma_S} = 
\sqrt{\frac{\left(\sigma_{S}+\sigma_{\rm
 SM}\right)\epsilon+B}{\sigma^2_{S}\epsilon^2{\cal L}}
+\frac{\delta\sigma_{\rm
 SM}^2\epsilon^2+(\delta{B})^2}{\sigma_{S}^2\epsilon^2}},
\label{eq:acc}
\end{align}
where $\delta{B}$ denotes the systematic uncertainty of the background
rate.
We take $\delta{B}=0.05B$, which may be achieved at the later stage of
the LHC experiment.
By solving Eq.~(\ref{eq:acc}), we obtain that $\sigma_S$ has to be
larger than 25~fb (63~fb) to achieve $\delta\sigma_S/\sigma_S<0.5$ (0.2)
with the integrated luminosity of ${\mathcal L}=300$~fb$^{-1}$. 
In our setup, the total uncertainty is dominated by the systematic
uncertainty from the background.
To reduce the statistical uncertainty smaller than the systematical one,
one needs only more than 10~fb$^{-1}$ of the data.
Thus, the accuracy will not be improved by accumulating the integrated
luminosity up to 3000~fb$^{-1}$, but is limited by the systematical
errors.
Increased integrated luminosity is, on the other hand, useful to reduce
the systematical uncertainty in the backgrounds. 
If we change our input by $B=7.2$~fb $\to3.6$~fb or
$\epsilon=0.03\to0.06$, the resulting values are modified to 14~fb
(36~fb) for $\delta\sigma_S/\sigma_S<0.5$ (0.2), respectively.
In the ideal case of $\delta B=0$, the parameter regions with
$\sigma_S\ge8$~fb (20~fb) can be observed at the $2\sigma$ ($5\sigma$)
CL.
To obtain better sensitivity, a considerable amount of efforts to reduce
the systematical uncertainties is required.

\begin{figure}[t]
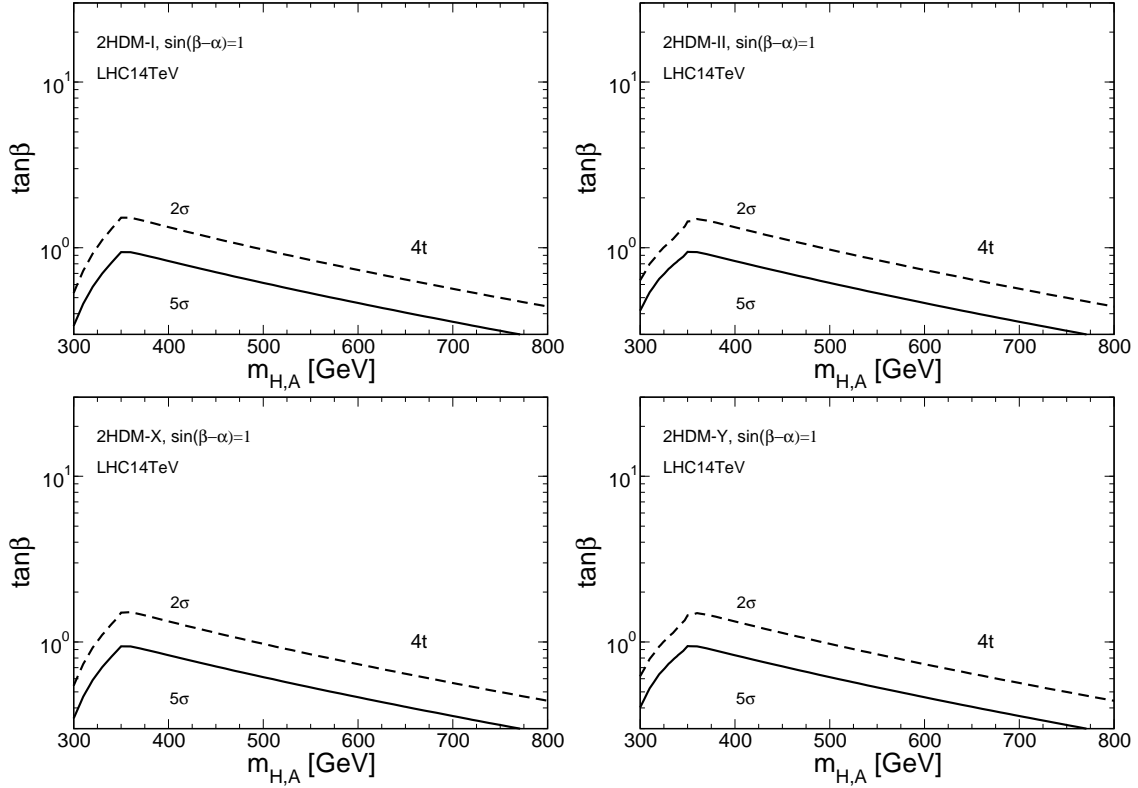

 \begin{center}
  \includegraphics[width=.45\textwidth,clip]{LHC4top_I.eps}
  \includegraphics[width=.45\textwidth,clip]{LHC4top_II.eps}
  \includegraphics[width=.45\textwidth,clip]{LHC4top_X.eps}
  \includegraphics[width=.45\textwidth,clip]{LHC4top_Y.eps}
 \end{center}
 \caption{Contour plot for the discovery reach of four top-quarks
 events at the LHC Run-II in the 2HDM.
 Upper limits on $\tan\beta$ for the discovery regions at the $2\sigma$
 [$5\sigma$] CL are shown in dashed [solid] lines.}\label{fig:LHC4top}
\end{figure}

In Fig.~\ref{fig:LHC4top}, we show the parameter regions in the 2HDM
with four types of Yukawa interactions where the additional Higgs bosons
contribution can be detected at the LHC in the four top-quarks events at
the $2\sigma$ and $5\sigma$ CL for the above setup.
It turned out that the dependence on the type of Yukawa interactions is
small, and only $\tan\beta\lesssim1.5$ can be probed at the $2\sigma$ CL
at most. 
Since these regions are constrained by flavor
experiments~\cite{Mahmoudi:2009zx,Cheng:2014ova}, the LHC searches in 4
top-quarks events may not have significant impact on exploring the
parameter regions in the 2HDMs.

%

\section{Multi-top-quarks production at the ILC}\label{sec:ilc}

In this section, we consider the four top-quarks production at the ILC.
In the 2HDM, the four top-quarks final state is generated via the pair and
single production of $H$ and/or $A$. 
For $\sqrt{s}>m_H+m_A$, pair production of $H$ and $A$, 
\begin{align}
 e^+e^-\to HA,
\end{align}
is kinematically possible and its cross section can be sizable.
The $HA$ pair production cross section does not depend on $\tan\beta$ at
the tree level.
Thus, the four top-quarks production rate depends on $\tan\beta$ only
through the decay branching ratio of $H$ and $A$. 

On the other hand, for $\sqrt{s}<m_H+m_A$, the pair production is
kinematically forbidden, but the single production process
\begin{align}
 e^+e^-\to t\bar tH (t\bar tA),\label{eq:ttha}
\end{align}
can still open as long as $\sqrt{s}>2m_t+m_{H(A)}$. 
The cross section of this process can be increased by the enhanced
Yukawa coupling of $H$ and $A$ to the top quarks.
Followed by the decays of $H$ and $A$ into $t\bar t$, four top-quarks
events occur from these processes.

Through the decay of top quarks, the signature of four top-quarks
production can be observed as all-hadronic, single lepton plus jets plus
missing momentum, dilepton plus jets plus missing, trilepton plus jets
plus missing, tetralepton plus jets plus missing channels.
Among dilepton plus jets plus missing channels, there are same-sign
and opposite-sign dilepton final states, where the former is expected to
have small backgrounds.
The branching fractions for these channels are listed in
Table~\ref{tab:bf}.
In the SM, the leading production mechanism of four top-quarks events is
 $e^+e^-\to t\bar{t}g^*\to t\bar{t}t\bar{t}$ via QCD interactions, thus
 the cross section is ${\mathcal O}(\alpha^2\alpha_s^2)$.
The next-to-leading production mechanism is full electroweak process,
 ${\mathcal O}(\alpha^4)$.
When we consider the final states including the decay of top quarks and
their detection at real experiments, there enter reducible backgrounds
via $e^+e^-\to t\bar t$, $t\bar t b\bar b$, $t\bar t \ell^+\ell^-$, etc. 
We estimate the contributions from these processes in the following
simulation analysis.

\begin{table}[tb]
 \begin{tabular}{l|l|l|l}
  \hline
  Decay modes & Final states & ${\mathcal R}$ (with $\tau$'s) &
  ${\mathcal R}$ (without $\tau$'s) \\
  \hline\hline
  all-hadron & $4j_b+8j$ &
	  $\left(\frac{2}{3}\right)^4\simeq0.2$ &
	      $\left(\frac{2}{3}\right)^4\simeq0.2$ \\
  single lepton + jets & $\ell^\pm + 4j_b+6j+\nu$ &
	  $\left(\frac{2}{3}\right)^3\cdot\frac{1}{3}\cdot4\simeq0.4$ & 
	      $\left(\frac{2}{3}\right)^3\cdot\frac{2}{9}\cdot4\simeq0.26$ \\
  S.S.\ dilepton + jets & $\ell^\pm\ell^\pm + 4j_b+4j +\nu\nu$ & 
	  $\left(\frac{2}{3}\right)^2\cdot
	  \left(\frac{1}{3}\right)^2\cdot2\simeq0.1$ &
	      $\left(\frac{2}{3}\right)^2\cdot
	      \left(\frac{2}{9}\right)^2\cdot2\simeq0.04$ \\
  O.S.\ dilepton + jets & $\ell^\pm\ell^\mp + 4j_b+4j +\nu\nu$
      & $\left(\frac{2}{3}\right)^2\cdot
	  \left(\frac{1}{3}\right)^2\cdot4\simeq0.2$
	  & $\left(\frac{2}{3}\right)^2\cdot
	      \left(\frac{2}{9}\right)^2\cdot4\simeq0.09$ \\
  trilepton + jets & $\ell^\pm\ell^\pm\ell^\mp + 4j_b+2j +\nu\nu\nu$ &
 	  $\;\frac{2}{3}\cdot\left(\frac{1}{3}\right)^3\cdot4\simeq0.1$ &
	      $\;\frac{2}{3}\cdot\left(\frac{2}{9}\right)^3\cdot4\simeq0.03$
	      \\
  tetralepton + jets & $\ell^+\ell^+\ell^-\ell^- + 4j_b +
      \nu\nu\nu\nu$ &
  	  $\left(\frac{1}{3}\right)^4\simeq0.01$ &
	      $\left(\frac{2}{9}\right)^4\simeq2.4\times10^{-3}$ \\
  \hline
 \end{tabular}
 \caption{Branching fractions of the four top-quarks signature in the
 decays of top quarks.
 Branching fractions where $\ell^\pm$ includes $\tau^\pm$ or not are
 listed.}\label{tab:bf}
\end{table}

First, we describe our simulation analysis for the detection of the 4
top-quarks events at the ILC with $\sqrt{s}=1$~TeV.
We use {\tt MadGraph5}~\cite{Alwall:2011uj} and {\tt
Pythia6}~\cite{Sjostrand:2006za} for generation of signal and background
events, with {\tt Tauola}~\cite{Jadach:1993hs} for tau lepton decays.
We evaluate the signal process by including both the single and
pair production amplitudes coherently.
The SM contribution to the $e^+e^-\to t\bar t t\bar t$ process is
estimated to be very small, giving $\sigma_{\rm
tot}=3.8\times10^{-3}$~fb at $\sqrt{s}=1$~TeV. 
Thus, for the integrated luminosity of ${\cal L}=1$~ab$^{-1}$, only a
few events are expected to be produced in the SM.
The background processes considered in our analysis are
\begin{align}
 & e^+e^-\to t\bar t,\\
 & e^+e^-\to t\bar t b\bar b,\\
 & e^+e^-\to t\bar t \ell^+\ell^-.\\
 & e^+e^-\to t\bar t W^+ W^-. \label{eq:ttww}
\end{align}
The second process includes $e^+e^-\to t\bar t g^*(\to b\bar b)$,
$e^+e^-\to t\bar t h(\to b\bar b)$, $e^+e^-\to t\bar t
Z/\gamma^*(\to b\bar b)$, $e^+e^-\to tbW^*(\to tb)$, and
$e^+e^-\to W^{+*}W^{-*}\to t\bar t b\bar b$.
$\ell^\pm$ in the third process mean the sum of $e^\pm$, $\mu^\pm$ and
$\tau^\pm$.
The other background processes are negligible.

To analyze the gererated events, we follow the designed performance of
the ILC detectors~\cite{Behnke:2013lya}. 
We take all  detectable particles whose pseudo rapidity satisfies
$|\eta|<1.5$. 
For charged particles, we further require their transverse momentum
satisfies $p_T>0.3$~GeV.
Four momenta of those particles are smeared by using the Gaussian
distribution with $\sigma_{p_T}/p_T = 10^{-4}p_T + 10^{-3}$ for charged
particles, $\sigma_{E}/E = 0.4/\sqrt{E} + 0.02$ for neutral hadrons, and
$\sigma_{E}/E = 0.15/\sqrt{E} + 0.01$ for photons, where $p_T$ and $E$
are given in an unit of GeV.

Among those particles, we select isolated leptons, $e$ and $\mu$, whose
energy satisfies $E_{\rm cone}\le\sqrt{6(E_{\ell}-15)}$ where $E_{\rm
cone}$ and $E_{\ell}$ are given in an unit of GeV.
$E_{\rm cone}$ is the summation of energies of particles inside the cone
around the lepton defined as $\cos\theta_{\rm cone}\ge0.98$ except the
lepton itself~\cite{Yonamine:2011jg}. 

After removing the isolated leptons from the list of particles, we
perform a jet clustering by using the Durham
algorithm~\cite{Catani:1991hj} with fixed $Y_{\rm cut}=5\times10^{-4}$
with the help of {\tt Fastjet}~\cite{Cacciari:2011ma}.
Thus, the number of jets in an event is flexible according to the event
structure.
For each clustered jet, we perform a flavour tagging.
If a jet contains only photons, it is tagged as a {\it photon-jet}.
$B$-tagging is performed stochastically by using the decay history of
particles which is available in the Monte-Carlo simulation.
If a jet contains $B$-hadrons ($D$-hadrons) in the decay history of
constituent particles, we tag it as a {\it b-jet} randomly with a
probability of 80\% (10\%). 
A probability of mis-tagging a jet which does not contain $B$ or
$D$-hadrons as a b-jet is set to be 3\%.
These probabilities correspond to {\it loose tagging} criteria given in
Ref.~\cite{Behnke:2013lya}.
We found that this loose criteria works better than rather tight
criteria to collect more signal events in the circumstances of small
backgrounds.
In addition, a jet is tagged as a {\it tau-jet}, if it contains 1 or 3
charged tracks and satisfies $E_{\rm cone}/E_{jet}>0.95$ where $E_{\rm
cone}$ is the summation of energies inside the small cone around a
direction of jet three momenta with $R=0.15$.
The other jets are assumed to be {\it light-jets}. 
The number of leptons in an event is counted as $N_{\rm lep}=N_{e}^{\rm
iso}+N_{\mu}^{\rm iso}+N_{\tau_j}$, where $N^{\rm iso}_{e(\mu)}$ is the
number of isolated $e$ ($\mu$) and $N_{\tau_j}$ is the number of
tau-jets.
The number of jets in an event is counted as $N_{\rm jet} =
N_{lj}+N_{bj}$, where $N_{lj(bj)}$ is the number of light-jets (b-jets).

\begin{figure}[t]
 \begin{center}
  \includegraphics[width=\textwidth,clip]{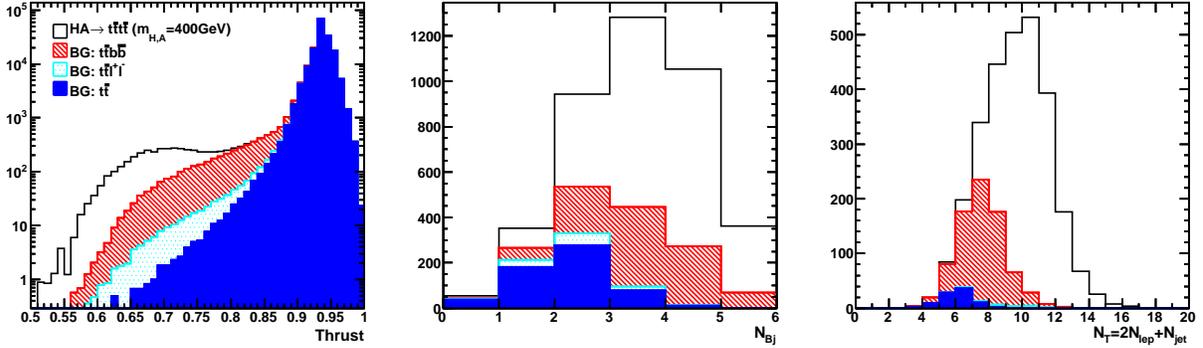}
  \caption{Kinematical distributions of signal and background events in
  Thrust (left), $N_{bj}$ (middle), and $N_T=2N_{\rm lep}+N_{\rm
  jet}$ (right) variables with loose $b$-tagging criteria.
  Signal events are displayed for $m_{H,A}=400$~GeV with a normalization
  of $\sigma_{\rm tot}=2.7$~fb which corresponds to $\tan\beta=1$ in
  Type-II 2HDM.}
  \label{fig:loose}
 \end{center}
\end{figure}

To extract the signal events out of the SM background, we impose following
selection cuts; (1) small thrust, $T<0.77$, (2) $N_{bj}\ge3$,
and (3) large ``{\it hard multiplicity}'', $N_T\equiv2N_{\rm lep}+N_{\rm
jet}\ge10$.
In Fig.~\ref{fig:loose}, we show distributions of signal and background
events in thrust, $N_{bj}$ and $N_T$ in the left, middle and right
panels, respectively. 
Backgrounds from $t\bar t$, $t\bar t \ell^+\ell^-$, and $t\bar t b\bar
b$ productions are shown in blue, cyan, and red histograms, respectively. 
Contributions from the $t\bar t W^+W^-$ process and the SM four top-quarks
production are omitted since these are small enough.
For the reference, signal events with $m_{H,A}=400$~GeV are shown on
top of them where the normalization is adjusted to the case of
$\tan\beta=1$ in the Type-II 2HDM.  
We find that large $t\bar t$ backgrounds are excluded by the thrust cut,
and a large amount of $t\bar t$, $t\bar t\ell^+\ell^-$ contributions is
excluded by the $N_{bj}$ cut. 
Furthermore, all the backgrounds are suppressed by applying the cut
on $N_T$. 

\begin{table}[t]
 \begin{tabular}{|l|c||c|c|c|}
  \hline
  Processes & Cross sections & \multicolumn{3}{c|} {Accumulated
  efficiencies} \\
  \hline
  & $\sigma_{\rm tot}$ [fb] & $T\le0.77$ & $N_{b_j}\ge 3$ &
  $2N_{\ell}+N_j\ge10$ \\
  \hline\hline
  \multicolumn{5}{|c|}{$e^+e^- (\to HA) \to t\bar t t\bar t$ [Type-II, $\sin(\beta-\alpha)=1$, $\tan\beta=1$]} \\
  \hline
  $m_{H,A}=360$~GeV & 4.3 & 71\% & 58\% & 34\% \\
  $\hphantom{m_{H,A}= } 400$~GeV & 2.7 & 92\% & 74\% & 43\% \\
  $\hphantom{m_{H,A}= } 440$~GeV & 1.3 & 96\% & 79\% & 47\% \\
  $\hphantom{m_{H,A}= } 480$~GeV & 0.30 & 96\% & 77\% & 46\% \\
  $\hphantom{m_{H,A}= } 500$~GeV & $7.5\times10^{-2}$ & 95\% & 77\% & 45\% \\
  $\hphantom{m_{H,A}= } 520$~GeV & $3.2\times10^{-2}$ & 95\% & 77\% & 45\% \\
  $\hphantom{m_{H,A}= } 560$~GeV & $1.0\times10^{-2}$ & 93\% & 76\% & 44\% \\
  \hline\hline
  $e^+e^-\to t\bar t$ & 166. & $3.6\times10^{-3}$ &
  $5.6\times10^{-4}$ &
  $2.0\times10^{-6}$ \\
  $e^+e^-\to t\bar t b\bar b$ & 5.0 & 19\% & 14\% & 0.66\% \\
  $e^+e^-\to t\bar t \ell^+\ell^-$ & 0.76 & 23\% & 3.8\% &
  1.4\% \\
  $e^+e^-\to t\bar t W^+W^-$ & 0.14 & 55\% & 15\% &
  3.0\% \\
  $e^+e^-\to t\bar t t\bar t$~(SM) & $3.8\times10^{-3}$ & 93\% & 74\% &
  41\% \\
  \hline
 \end{tabular}
\caption{The total cross section and the accumulated efficiencies by
 kinematical cuts for the signal and background processes at the ILC
 with $\sqrt{s}=1$~TeV.
 Signal cross sections and efficiencies are calculated for
 $m_{H,A}=360$~GeV to 560~GeV with $\tan\beta=1$ for Type-II as a
 reference.}
\label{tab:cut}
\end{table}

The background reduction and signal detection efficiencies are
summarized in Table~\ref{tab:cut} where the background
process Eq.~(\ref{eq:ttww}) and the SM four top-quarks production
processes are also included for the reference.
In our simulation, the background reduction rates by the above three cuts
are ${\cal O}(10^{-6})$ for $t\bar t$, 0.66\% for $t\bar t b\bar b$,
1.4\% for $t\bar t\ell^+\ell^-$, and 3\% for $t\bar t W^+W^-$.
With the integrated luminosity of ${\cal L}=1$~ab$^{-1}$, only around
47.8 events are expected to be observed.
From the SM four top-quarks production, we expect 1.6 events.
On the other hand, for the signal process of additional Higgs bosons,
around 34\% to 47\% of events are remained after these cuts, depending
on the mass.

By using the signal detection efficiency $\epsilon_S$ with the
expected number of background rate $B=49.4$~ab at ${\cal
L}=1$~ab$^{-1}$, we can estimate the minimum value of the total cross
section for the signal process to be identified in a certain accuracy.
We take into account only the statistical uncertainties due to the
presence of backgrounds, since at lepton colliders
systematical uncertainties can be expected to be well under control.
Thus, the uncertainty of observing the total cross section of signal
events is estimated as
\begin{align}
\frac{\delta\sigma_{S}}{\sigma_S}=
\sqrt{\frac{\sigma_S\epsilon_S+B}{\sigma^2_S\epsilon^2_S{\cal
 L}}}.
\end{align} 
We find that the signal process can be detected at the $2\sigma$
($5\sigma$) CL if the total cross section is above 0.034-0.048~fb
(0.11-0.15~fb), depending on the mass of additional Higgs bosons through
the signal detection efficiency.

\begin{figure}[t]
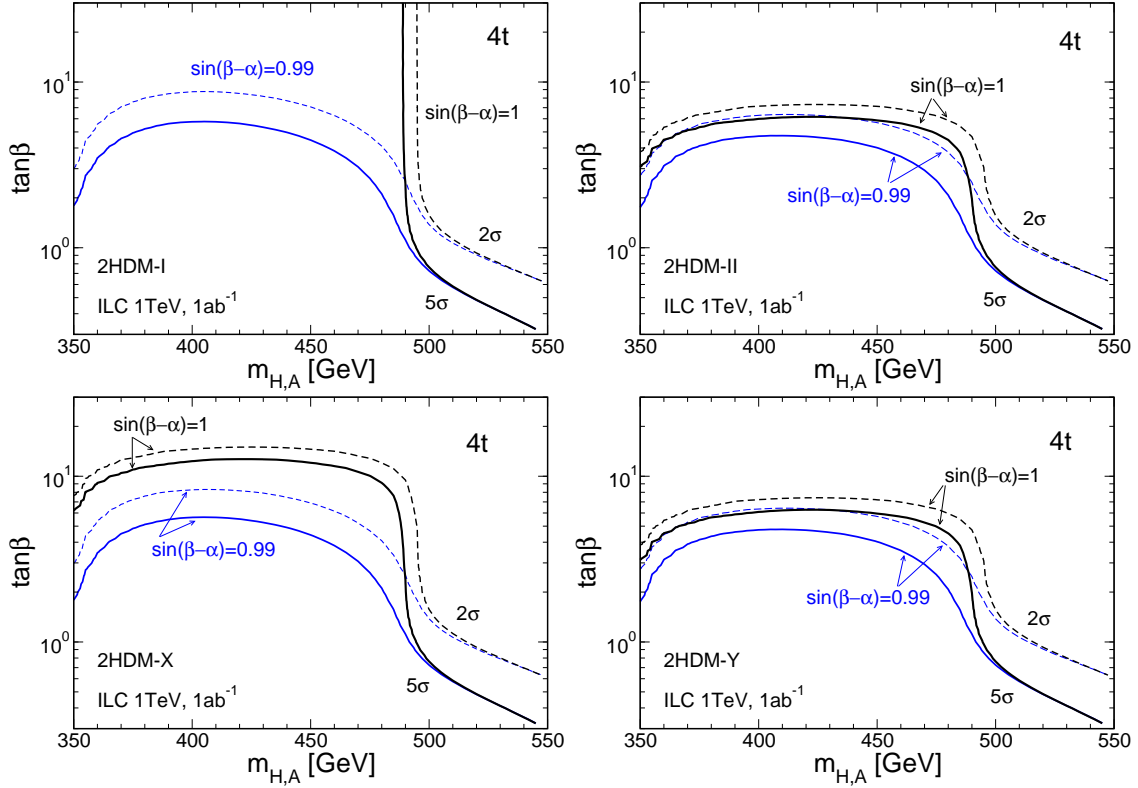

 \begin{center}
  \includegraphics[width=.45\textwidth,clip]{ILC4top_I.eps}
  \includegraphics[width=.45\textwidth,clip]{ILC4top_II.eps}
  \includegraphics[width=.45\textwidth,clip]{ILC4top_X.eps}
  \includegraphics[width=.45\textwidth,clip]{ILC4top_Y.eps}
 \end{center}
 \caption{Contour plot of the discovery reach of the four top-quarks
 production in the 2HDMs at the ILC $\sqrt{s}=1$~TeV with ${\cal
 L}=1$~ab$^{-1}$ data. 
 Upper limits on $\tan\beta$ for the discovery regions at the $2\sigma$
 [$5\sigma$] CL are shown in solid [dashed] lines. 
The similar results but for $\sin(\beta-\alpha)=0.99$ are also shown
 in blue lines. 
}\label{fig:ILC4top}
\end{figure}

In Fig.~\ref{fig:ILC4top}, we evaluate the parameter regions in the
$(m_{H,A}, \tan\beta)$ plane in the 2HDMs where the four top-quarks
events can be detected at the ILC. 
Solid (dashed) contours show the regions where the signal can be
 detected at the $2\sigma$ ($5\sigma$) CL.
For Type-II and Type-Y 2HDM, $\tan\beta$ up to around 7 (6) can be
probed for the mass of additional Higgs bosons up to $m_{H,A}\lesssim
500$~GeV.
Above $m_{H,A}\simeq 500$~GeV, the signal can be produced by using the
 single Higgs production processes, Eq.~(\ref{eq:ttha}), but only
 visible for small  $\tan\beta$ below unity.
For Type-X 2HDM, $\tan\beta$ up to around 15 (12) can be probed for
 $m_{H,A}\lesssim 500$~GeV, because of the fact that the $H/A\to t\bar
 t$ decay branching rate remains to be dominant up to larger $\tan\beta$
 than that in Type-II and Y. 
For Type-I 2HDM, since the $H/A\to t\bar t$ decay mode dominates for any
 value of $\tan\beta$, the four top-quarks events can be observed for any
 value of $\tan\beta$, as long as $m_{H,A}\lesssim 500$~GeV. 

We note that, however, the result for Type-I is due to the fact
that we take the SM-like limit, $\sin(\beta-\alpha)=1$, where no $H\to
WW$, $ZZ$, $hh$ or $A\to Zh$ decay is induced.
For example, if we take $\sin(\beta-\alpha)=0.99$, these decay modes
become non-zero, and even dominant for larger
$\tan\beta$~\cite{Kanemura:2014bqa}. 
The value of $\sin(\beta-\alpha)=0.99$ can be detected by the precision
measurement of $hVV$ coupling constants at the ILC with
$\sqrt{s}=500$~GeV and ${\cal L}=1.6$~ab$^{-1}$ at the $5\sigma$
CL~\cite{Asner:2013psa}. 
In this case, the four top-quarks events can be observed up to about
$\tan\beta\simeq 8$~(5) at the $2\sigma$ ($5\sigma$) CL for the four
types of Yukawa models. 
The discovery reaches for $\sin(\beta-\alpha)=0.99$ are also shown in
each figure in Fig.~\ref{fig:ILC4top} in blue lines. 

We now compare the impact of the searches for multi-top-quarks events
at the LHC and the ILC on exploring the parameter regions in the 2HDM.
At the LHC Run-II, by the searches for four top-quarks events, the
parameter regions with $\tan\beta\lesssim1.5$ can be surveyed in all
types of Yukawa interactions. 
Because the experimental uncertainties are dominated by the systematical
ones, the results will not be improved at the LHC 3000~fb$^{-1}$. 
On the other hand, at the ILC 1~TeV, we find that the parameter regions
with larger $\tan\beta$ can be surveyed as long as
$m_{H,A}\lesssim500$~GeV. 
For Type-II and Type-Y (Type-X), the parameter regions with
$\tan\beta\simeq 7$ (15) can be explored. 
For Type-I, because the decay mode into a top-quark
pair is dominant for any value of $\tan\beta$, detection of four top-quarks
events is anticipated in any value of $\tan\beta$. 
This result is modified to $\tan\beta\lesssim9$ in the case of
$\sin(\beta-\alpha)=0.99$.
Therefore, the searches for four top-quarks events can explore wider
parameter regions at the ILC 1~TeV than at the LHC 3000~fb$^{-1}$.

\section{Discussions}\label{sec:dis}

In this section, we discuss further investigation on the
multi-top-quarks events at the LHC and the ILC, regarding the mass
determination of the additional Higgs bosons and the discrimination of
the type of Yukawa interactions. 
In addition, we give some comments on the searches for multi-top-quarks
events at future multi-TeV lepton collider, CLIC.

First, we discuss the mass reconstruction of additional Higgs bosons in
the decay mode into a top-quark pair.
At the LHC, even if some excess in the events for the lepton plus
multi-jets channel is observed, the mass reconstruction of additional
Higgs boson is difficult, since one has to form an invariant mass of six
jets in the pretense of more jets which come from the top quarks and
the ISR/FSR jets.
The combinatorial uncertainty and the jet energy resolution prevent us
from constructing a clear peak in the invariant mass distribution. 
At the ILC, the mass reconstruction by kinematical methods is still
difficult, because of the combinatorial uncertainty and the fact that
the top quarks from the decay of additional Higgs bosons with the mass
of 350-500~GeV have relatively low velocity.
Thus, for the mass determination, we may rely on the other decay modes,
accompanied to the $t\bar t$ mode, such as $b\bar b$ and
$\tau^+\tau^-$~\cite{Kanemura:2011kx}.

Since the enhancement of the couplings of $H$ and $A$ to top quarks
is true for the all four types of Yukawa interactions, the
discrimination of the type of Yukawa interactions needs to see the other
decay modes, such as $b\bar{b}$ for Type-II and Type-Y and
$\tau^+\tau^-$ for Type-II and Type-X.
For Type-I, the other decay modes are all suppressed.
Therefore, non-observation of the other fermionic channel may be the
signature of Type-I.

For the above two reasons, combinations of the searches of additional
Higgs bosons in different decay modes are important.
Depending on the type of Yukawa interactions and the value of
$\tan\beta$, the decay modes into $b\bar b$ and $\tau^+\tau^-$ can be
explored at the LHC and/or the ILC~\cite{Kanemura:2014dea}. 
By observing several decay modes and their event rates, $\tan\beta$ can
be determined experimentally~\cite{Kanemura:2013eja}.
In the case of $\sin(\beta-\alpha)<1$, the decay modes of $H\to hh$,
$WW$, $ZZ$, and $A\to Zh$ can be sizable, especially for Type-I.  
By the combinations of the measurements of these modes, experimental
determination of $\sin(\beta-\alpha)$ may be performed.

Finally, we give some comments on the searches at the
CLIC~\cite{Linssen:2012hp}. 
At the CLIC, the collision energy of multi-TeV, such as 3~TeV, is
proposed. 
With $\sqrt{s}=3$~TeV, the mass reach would be extended up to about
1.5~TeV.
This is totally above the scope at the LHC.
Thus, the direct searches at the CLIC would have a great impact in any
decay mode of additional Higgs bosons.
The top quarks from the decay of such heavier Higgs bosons are more
energetic so that the mass reconstruction by using boosted top-jet 
measurement can be realistic.
In the case that $H$ and $A$ are produced with large velocity, the decay
products of each Higgs boson are well separated into different
hemispheres.
Therefore, the mass of additional Higgs boson can be reconstructed by
using the invariant mass of all decay products in one
hemisphere~\cite{Battaglia:2010xq}.

\section{Conclusions}\label{sec:conc}

We have studied the direct searches of the additional Higgs bosons in 
multi-top-quarks events at the future LHC and the ILC experiments. 
Additional Higgs bosons are predicted in any kind of models with
extended Higgs sectors, and the detection of them at colliders
is a clear signature of the physics beyond the standard model.
As a benchmark model of the extended Higgs sector, we have considered
 the two Higgs doublet models with discrete symmetry with four types of
 Yukawa interactions. 

At the LHC, the signals of four top-quarks events suffer from
systematical uncertainties of estimating the SM backgrounds. 
Therefore, the searches can only survey the parameter regions with 
 $\tan\beta\lesssim1.5$ which is disfavored by experimental constraints
 from flavor physics without adding new particle contents to the model.
At the ILC, although the mass reach is almost limited by its beam
energy, we have shown that the parameter regions with
$\tan\beta\lesssim8$-15 can be surveyed depending on the type of Yukawa
interactions. 
We have also discussed that further investigation may be
 performed at the ILC, such as discrimination of the type of Yukawa
 interactions and determination of the mass of additional Higgs bosons
 by combining the observations of the other decay modes of $H$ and $A$ in
 addition to the $t\bar t$ mode.

\acknowledgments
H.Y.\ thanks Quantum Universe Center at KIAS for the warm hospitality.
This work was supported, in part, by Grant-in-Aid for Scientific research
from the Ministry of Education, Science, Sports, and Culture (MEXT),
Japan, Nos.\ 22244031, 23104006 and 24340046, and NSC of ROC. 




\begin{thebibliography}{99}

\bibitem{Aad:2012tfa}
  G.~Aad {\it et al.}  [ATLAS Collaboration],
  Phys.\ Lett.\ B {\bf 716} (2012) 1.

\bibitem{Chatrchyan:2012ufa}
  S.~Chatrchyan {\it et al.}  [CMS Collaboration],
  Phys.\ Lett.\ B {\bf 716} (2012) 30.

\bibitem{Aad:2013wqa}
  G.~Aad {\it et al.}  [ATLAS Collaboration],
  Phys.\ Lett.\ B {\bf 726} (2013) 88.

\bibitem{Aad:2013xqa}
  G.~Aad {\it et al.}  [ATLAS Colaboration],
  Phys.\ Lett.\ B {\bf 726} (2013) 120.

\bibitem{Chatrchyan:2013iaa}
  S.~Chatrchyan {\it et al.}  [CMS Collaboration],
  JHEP {\bf 1401} (2014) 096.

\bibitem{Chatrchyan:2013mxa}
  S.~Chatrchyan {\it et al.}  [CMS Collaboration],
  Phys.\ Rev.\ D {\bf 89} (2014) 9,  092007.

\bibitem{Aad:2015zhl} 
  G.~Aad {\it et al.}  [ATLAS and CMS Collaborations],
  arXiv:1503.07589 [hep-ex].



\bibitem{Asner:2013psa} 
  D.~M.~Asner, T.~Barklow, C.~Calancha, K.~Fujii, N.~Graf, H.~E.~Haber, A.~Ishikawa and S.~Kanemura {\it et al.},
  arXiv:1310.0763 [hep-ph].

\bibitem{Dawson:2013bba}
  S.~Dawson, A.~Gritsan, H.~Logan, J.~Qian, C.~Tully, R.~Van Kooten, A.~Ajaib and A.~Anastassov {\it et al.},
  arXiv:1310.8361 [hep-ex].



\bibitem{Djouadi:2007ik} 
  G.~Aarons {\it et al.}  [ILC Collaboration],
  arXiv:0709.1893 [hep-ph].

\bibitem{Behnke:2013lya}
T.~Behnke {\it et al.},
arXiv:1306.6329 [physics.ins-det].

\bibitem{Gomez-Ceballos:2013zzn} 
  M.~Bicer {\it et al.}  [TLEP Design Study Working Group Collaboration],
  JHEP {\bf 1401} (2014) 164.

\bibitem{Linssen:2012hp} 
  L.~Linssen, A.~Miyamoto, M.~Stanitzki and H.~Weerts,
  arXiv:1202.5940 [physics.ins-det].


\bibitem{Kanemura:2014bqa} 
  S.~Kanemura, K.~Tsumura, K.~Yagyu and H.~Yokoya,
  Phys.\ Rev.\ D {\bf 90} (2014) 075001.

\bibitem{Kanemura:2015mxa}
  S.~Kanemura, M.~Kikuchi and K.~Yagyu,
  Nucl.\ Phys.\ B {\bf 896} (2015) 80.


\bibitem{Agashe:2014kda} 
  K.~A.~Olive {\it et al.}  [Particle Data Group Collaboration],
  Chin.\ Phys.\ C {\bf 38} (2014) 090001.


\bibitem{Kanemura:2014dea} 
  S.~Kanemura, H.~Yokoya and Y.~J.~Zheng,
  Nucl.\ Phys.\ B {\bf 886} (2014) 524.


	
\bibitem{Gunion:1989we} 
  J.~F.~Gunion, H.~E.~Haber, G.~L.~Kane and S.~Dawson,
  Front.\ Phys.\  {\bf 80} (2000) 1.

\bibitem{Branco:2011iw}
  G.~C.~Branco {\it et al.},
  Phys.\ Rept.\  {\bf 516} (2012) 1.



\bibitem{Cheung:1995eq}
  K.~-m.~Cheung,
  hep-ph/9507411.

\bibitem{Spira:1997ce}
  M.~Spira and J.~D.~Wells,
  Nucl.\ Phys.\ B {\bf 523} (1998) 3.

\bibitem{Lillie:2007hd} 
  B.~Lillie, J.~Shu and T.~M.~P.~Tait,
  JHEP {\bf 0804} (2008) 087.

\bibitem{Chen:2008hh} 
  C.~R.~Chen, W.~Klemm, V.~Rentala and K.~Wang,
  Phys.\ Rev.\ D {\bf 79} (2009) 054002.

\bibitem{Barger:2010uw} 
  V.~Barger, W.~Y.~Keung and B.~Yencho,
  Phys.\ Lett.\ B {\bf 687} (2010) 70.

\bibitem{Jung:2010ms}
  S.~Jung and J.~D.~Wells,
  JHEP {\bf 1011} (2010) 001.

\bibitem{Cacciapaglia:2011kz}
  G.~Cacciapaglia, R.~Chierici, A.~Deandrea, L.~Panizzi, S.~Perries and S.~Tosi,
  JHEP {\bf 1110} (2011) 042.

\bibitem{Gregoire:2011ka} 
  T.~Gregoire, E.~Katz and V.~Sanz,
  Phys.\ Rev.\ D {\bf 85} (2012) 055024.

\bibitem{AguilarSaavedra:2011ck} 
  J.~A.~Aguilar-Saavedra and J.~Santiago,
  Phys.\ Rev.\ D {\bf 85} (2012) 034021.

\bibitem{Han:2012qu} 
  C.~Han, N.~Liu, L.~Wu and J.~M.~Yang,
  Phys.\ Lett.\ B {\bf 714} (2012) 295.

\bibitem{Chen:2014ewl} 
  C.~R.~Chen,
  Phys.\ Lett.\ B {\bf 736} (2014) 321.

\bibitem{Dev:2014yca} 
  P.~S.~B.~Dev and A.~Pilaftsis,
  JHEP {\bf 1412} (2014) 024.

\bibitem{Greiner:2014qna} 
  N.~Greiner, K.~Kong, J.~C.~Park, S.~C.~Park and J.~C.~Winter,
  JHEP {\bf 1504} (2015) 029.

\bibitem{Beck:2015cga} 
  L.~Beck, F.~Blekman, D.~Dobur, B.~Fuks, J.~Keaveney and K.~Mawatari,
  arXiv:1501.07580 [hep-ph].

\bibitem{Craig:2015jba} 
  N.~Craig, F.~D'Eramo, P.~Draper, S.~Thomas and H.~Zhang,
  arXiv:1504.04630 [hep-ph].


\bibitem{Djouadi:1996ah}
  A.~Djouadi, H.~E.~Haber and P.~M.~Zerwas,
  Phys.\ Lett.\ B {\bf 375} (1996) 203.

\bibitem{Gunion:1988tf}
  J.~F.~Gunion, L.~Roszkowski, A.~Turski, H.~E.~Haber, G.~Gamberini, B.~Kayser, S.~F.~Novaes and F.~I.~Olness {\it et al.},
  Phys.\ Rev.\ D {\bf 38} (1988) 3444.



\bibitem{Borzumati:1999th}
  F.~Borzumati, J.~-L.~Kneur and N.~Polonsky,
  Phys.\ Rev.\ D {\bf 60} (1999) 115011.

\bibitem{Plehn:2002vy}
  T.~Plehn,
  Phys.\ Rev.\ D {\bf 67} (2003) 014018;

  E.~L.~Berger, T.~Han, J.~Jiang and T.~Plehn,
  Phys.\ Rev.\ D {\bf 71} (2005) 115012.


\bibitem{Kanemura:2000cw}
  S.~Kanemura, S.~Moretti and K.~Odagiri,
  JHEP {\bf 0102} (2001) 011.

\bibitem{Moretti:2002pa}
  S.~Moretti,
  Eur.\ Phys.\ J.\ direct C {\bf 4} (2002) 15.

\bibitem{Kiyoura:2003tg} 
  S.~Kiyoura, S.~Kanemura, K.~Odagiri, Y.~Okada, E.~Senaha, S.~Yamashita and Y.~Yasui,
  hep-ph/0301172.







\bibitem{Barger:1989fj}
  V.~D.~Barger, J.~L.~Hewett and R.~J.~N.~Phillips,
  Phys.\ Rev.\ D {\bf 41} (1990) 3421.

\bibitem{Grossman:1994jb}
  Y.~Grossman,
  Nucl.\ Phys.\ B {\bf 426} (1994) 355.

\bibitem{Aoki:2009ha}
  M.~Aoki, S.~Kanemura, K.~Tsumura and K.~Yagyu,
  Phys.\ Rev.\ D {\bf 80} (2009) 015017.


\bibitem{Glashow:1976nt} 
  S.~L.~Glashow and S.~Weinberg,
  Phys.\ Rev.\ D {\bf 15} (1977) 1958.



\bibitem{Kanemura:1993hm} 
  S.~Kanemura, T.~Kubota and E.~Takasugi,
  Phys.\ Lett.\ B {\bf 313} (1993) 155.

\bibitem{Akeroyd:2000wc} 
  A.~G.~Akeroyd, A.~Arhrib and E.~-M.~Naimi,
  Phys.\ Lett.\ B {\bf 490} (2000) 119.

\bibitem{Ginzburg:2005dt} 
  I.~F.~Ginzburg and I.~P.~Ivanov,
  Phys.\ Rev.\ D {\bf 72}0 (2005) 11501.


\bibitem{Deshpande:1977rw} 
  N.~G.~Deshpande and E.~Ma,
  Phys.\ Rev.\ D {\bf 18} (1978) 2574.

\bibitem{Kanemura:1999xf}
  S.~Kanemura, T.~Kasai and Y.~Okada,
  Phys.\ Lett.\ B {\bf 471} (1999) 182.

\bibitem{Nie:1998yn}
  S.~Nie and M.~Sher,
  Phys.\ Lett.\ B {\bf 449} (1999) 89.



\bibitem{Gunion:2002zf}
  J.~F.~Gunion and H.~E.~Haber,
  Phys.\ Rev.\ D {\bf 67} (2003) 075019.


\bibitem{Kanemura:2001hz}
  S.~Kanemura and C.~P.~Yuan,
  Phys.\ Lett.\ B {\bf 530} (2002) 188.

\bibitem{Cao:2003tr}
  Q.~H.~Cao, S.~Kanemura and C.~P.~Yuan,
  Phys.\ Rev.\ D {\bf 69} (2004) 075008.


\bibitem{Djouadi}
  A.~Djouadi,
  Phys.\ Rept.\  {\bf 457} (2008) 1; 
  Phys.\ Rept.\  {\bf 459} (2008) 1. 

\bibitem{Alwall:2011uj}
  J.~Alwall, M.~Herquet, F.~Maltoni, O.~Mattelaer and T.~Stelzer,
  JHEP {\bf 1106} (2011) 128.

\bibitem{CTEQ6L} 
  J.~Pumplin, D.~R.~Stump, J.~Huston, H.~L.~Lai, P.~M.~Nadolsky and W.~K.~Tung,
  JHEP {\bf 0207} (2002) 012.


\bibitem{TheATLAScollaboration:2013jha} 
  The ATLAS collaboration,
  ATLAS-CONF-2013-051, ATLAS-COM-CONF-2013-055.


\bibitem{Khachatryan:2014sca} 
  V.~Khachatryan {\it et al.}  [CMS Collaboration],
  JHEP {\bf 1411} (2014) 154.



\bibitem{Bevilacqua:2012em} 
  G.~Bevilacqua and M.~Worek,
  JHEP {\bf 1207} (2012) 111.





\bibitem{Mahmoudi:2009zx} 
  F.~Mahmoudi and O.~Stal,
  Phys.\ Rev.\ D {\bf 81} (2010) 035016.

\bibitem{Cheng:2014ova} 
  X.~D.~Cheng, Y.~D.~Yang and X.~B.~Yuan,
  Eur.\ Phys.\ J.\ C {\bf 74}, no. 10 (2014) 3081.




\bibitem{Sjostrand:2006za} 
  T.~Sjostrand, S.~Mrenna and P.~Z.~Skands,
  JHEP {\bf 0605} (2006) 026.

\bibitem{Jadach:1993hs} 
  S.~Jadach, Z.~Was, R.~Decker and J.~H.~Kuhn,
  Comput.\ Phys.\ Commun.\  {\bf 76} (1993) 361.


\bibitem{Yonamine:2011jg} 
  R.~Yonamine, K.~Ikematsu, T.~Tanabe, K.~Fujii, Y.~Kiyo, Y.~Sumino and H.~Yokoya,
  Phys.\ Rev.\ D {\bf 84} (2011) 014033.

\bibitem{Catani:1991hj} 
  S.~Catani, Y.~L.~Dokshitzer, M.~Olsson, G.~Turnock and B.~R.~Webber,
  Phys.\ Lett.\ B {\bf 269} (1991) 432.

\bibitem{Cacciari:2011ma} 
  M.~Cacciari, G.~P.~Salam and G.~Soyez,
  Eur.\ Phys.\ J.\ C {\bf 72} (2012) 1896.


\bibitem{Kanemura:2011kx}
  S.~Kanemura, K.~Tsumura and H.~Yokoya,
  Phys.\ Rev.\ D {\bf 85} (2012) 095001.

\bibitem{Kanemura:2013eja}
  S.~Kanemura, K.~Tsumura and H.~Yokoya,
  Phys.\ Rev.\ D {\bf 88} (2013) 055010.


\bibitem{Battaglia:2010xq} 
  M.~Battaglia and G.~Servant,
  Nuovo Cim.\ C {\bf 033N2} (2010) 203.


\end{thebibliography}
\end{document}